\documentclass[aip,pop,reprint]{revtex4-1}

\usepackage{amsmath,amsthm,amssymb,graphicx}
\usepackage{color}

\draft % marks overfull lines with a black rule on the right

\newtheorem{thm}{Theorem}

\begin{document}




%\preprint{}

\title{A Generalized Flux Function for Three-dimensional Magnetic Reconnection} %Title of paper








\author{A. R. Yeates}

\email[]{anthony@maths.dundee.ac.uk}


%\thanks{}


\author{G. Hornig}

\email[]{gunnar@maths.dundee.ac.uk}

\affiliation{Division of Mathematics, University of Dundee, Dundee DD1 4HN, UK}





\date{\today}

\begin{abstract}
The definition and measurement of magnetic reconnection in three-dimensional magnetic fields with multiple reconnection sites is a challenging problem, particularly in fields lacking null points. We propose a generalization of the familiar two-dimensional concept of a magnetic flux function to the case of a three-dimensional field connecting two planar boundaries. Using hyperbolic fixed points of the field line mapping, and their global stable and unstable manifolds, we define a unique flux partition of the magnetic field. This partition is more complicated than the corresponding (well-known) construction in a two-dimensional field, owing to the possibility of heteroclinic points and chaotic magnetic regions. Nevertheless, we show how the partition reconnection rate is readily measured with the generalized flux function. We relate our partition reconnection rate to the common definition of three-dimensional reconnection in terms of integrated parallel electric field. An analytical example demonstrates the theory, and shows how the flux partition responds to an isolated reconnection event.
\end{abstract}


\maketitle %\maketitle must follow title, authors, abstract and \pacs



%%%%%%%%%%%%%%%%%%%%
%%%%%%%%%%%%%%%%%%%%
\section{Introduction}

This paper presents a new method for measuring magnetic reconnection in a three-dimensional (3D) magnetic field. Reconnection is a fundamental physical process in any highly-conducting plasma, yet remains poorly understood owing to the challenging range of lengthscales involved \citep{priest2000,birn2007}. In 3D magnetic fields, progress is hampered by the difficulty in defining and measuring reconnected flux, particularly if there are multiple interacting reconnection sites. It is this problem that we seek to address.

There are two contrasting ways to measure reconnection rates in 3D. The first uses the parallel electric field integrated along magnetic field lines\citep{schindler1988}, while the second counts the transfer of flux between distinct flux domains. Here we pursue the second approach, where the task is twofold: to define a partition of the flux and to measuring the rate of transfer between fluxes in this partition. We call this a reconnection rate with respect to a partition, or a  \emph{partition reconnection rate}, to distinguish it from the first case. Such a partition reconnection rate can capture only reconnection processes that change fluxes between the partition domains, and not those within any individual flux domain. However, this rate is in many applications the most relevant information, detemining the stability and  dynamics of the system.

In a two-dimensional (2D) field ${\bf B}=B_x(x,y){\bf e}_x + B_y(x,y){\bf e}_y$, there is a natural choice for such a partition and correspondingly for the reconnection rate: write ${\bf B}$ in terms of a \emph{flux function} $A(x,y)$ where ${\bf B}=\nabla\times A\,{\bf e}_z$. The different fluxes in the partition correspond to the regions of the plane bounded by separatrices, which are the global stable and unstable manifolds of hyperbolic nulls (x-points). The magnetic flux within each such region (per unit height in the ignorable $z$ direction) is measured by the difference in $A$ between appropriately chosen nulls. These fluxes are invariant under an ideal evolution, while in a non-ideal evolution the change in fluxes is measured exactly by the change in the values of $A$ at the (discrete) set of null points. This defines an unambiguous global reconnection rate which is readily computed even in turbulent 2D fields with many nulls \citep{servidio2009}.
 
In 3D, the situation is more complicated. Here a natural partition also arises from the existence of null points in the domain. The 2D invariant manifolds (fan surfaces) associated with these null points form a coarse but natural  partition of the flux\citep{longcope2009,haynes2010}. This inherent topological structure has been used successfully to quantify reconnection\citep{pontin2011a}. Nulls and separators in particular are favoured locations to detect global changes in connectivity because they are locations where distinct flux domains come into close proximity. 

Another possible flux partition arises in a field connected to a physical boundary, such as the photosphere of the Sun, where the sign of the normal field component divides the boundary into regions of positive and negative magnetic polarity. This boundary partition extends to a partition of magnetic flux connected to the boundary, by following the field lines from the different polarity regions into the volume\citep{longcope2005}.

There are, however, many examples of magnetic fields where either the above described partitions are too coarse or null points do not exist in the domain. Examples are a single coronal loop or the magnetic field in a tokamak. A  generic 3D magnetic field in this situation does not possess a foliation of flux surfaces. This ``worst case" is the situation which we want to investigate here.  More specifically, we assume a simply-connected domain in which  all field lines stretch between two planar boundaries. In magnetospheric reconnection studies  this is often referred to as the ``guide field'' case. We present a general way to define the flux partition in such a field, using distinguished hyperbolic orbits, and measure reconnection by introducing a generalized version of the 2D flux function $A$. Not only is this a natural topological flux partition when there are no magnetic null points, but it retains the simplicity of the 2D method when it comes to measuring the partition reconnection rate.

As may be expected, there are some complexities that do not arise in the 2D case. Chief among these is the possibility of chaos in the field line mapping. This is well-known from the study of area-preserving mappings as models for the magnetic field in toroidal fusion devices \citep{morrison2000}. Indeed, recent results in tokamak experiments show heating on the vessel walls consistent with the breakdown of confinement and chaotic transport of magnetic flux through homoclinic tangles, as found in numerical simulations \citep{evans2005,borgogno2008,viana2011}. Though they were recognised by Poincar\'{e} in the 19th Century, it is only recently that detailed analysis of the structure of homoclinic tangles has been applied to measure and predict the transport of trajectories in these chaotic regions. A primary application has been 2D fluids with time-dependent velocity fields \citep{romkedar1990a,ryrie1992,beigie1994,mancho2006}. Here, we show how these ideas can be applied to define and measure a natural reconnection rate in 3D magnetic fields.

%%%%%%%%%%%%%%%%%%%%
%%%%%%%%%%%%%%%%%%%%
\section{Two-dimensional magnetic fields} \label{sec:2d}

\begin{figure}[htbp]
\begin{center}
\includegraphics[width=\columnwidth]{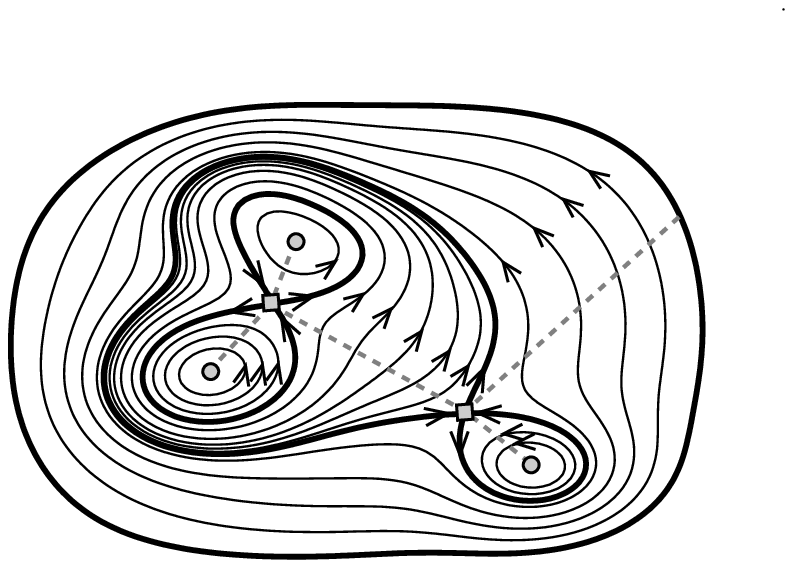}
\caption{A 2D magnetic field, showing nulls (hyperbolic as squares, elliptic as circles) and separatrices (thick lines). Dashed grey lines show the differences in $A$ that measure the partition fluxes.}
\label{fig:2d}
\end{center}
\end{figure}

We first review the basic properties of the flux function $A(x,y)$ of a 2D magnetic field ${\bf B}=\nabla\times A{\bf e}_z$ (Figure \ref{fig:2d}).

\begin{enumerate}
\item $A$ is constant along magnetic field lines (${\bf B}\cdot\nabla A=0$).
\item Consider two vertical lines through the points $(x_1,y_1)$ and $(x_2,y_2)$. The magnetic flux through any surface bounded by the two lines and the planes $z=0$, $z=1$ is $A(x_2,y_2) - A(x_1,y_1)$ (w.r.t.~the orientation of the surface). This is the reason for the name \emph{flux function}. It works because
\begin{equation}
A(x,y) = \int_0^1A(x,y) dz = \int_0^1{\bf A}\cdot\,d{\bf l},
\end{equation}
where ${\bf A}=A(x,y){\bf e}_z$ is a vector potential for ${\bf B}$.
\item For an ideal evolution 
\[
\frac{\partial{\bf B}(x,y,t)}{\partial t} - \nabla\times\big({\bf v}(x,y,t)\times{\bf B}(x,y,t)\big)=0,
\]
 $A(x,y)$ can be chosen as an ideal invariant,
\begin{equation}
\frac{\partial A}{\partial t} + {\bf v}\cdot\nabla A = 0.
\end{equation}
\end{enumerate}

A 2D magnetic field is naturally partitioned by the x-points (hyperbolic nulls) and their separatrices. The separatrices---shown by thick lines in Figure \ref{fig:2d}---are the topologically distinguished field lines given by the global stable and unstable manifolds of each x-point. These manifolds are tangent at the null to the unstable or stable eigenvectors of the local linearisation, and are uniquely defined by extending forwards or backwards along the flow. They are invariant subspaces of the field line flow (i.e., they are field lines) that delineate topologically distinct regions. Each region has a well-defined flux measured by the difference in $A$ between two (not necessarily unique) null points (joined by dashed lines in Figure \ref{fig:2d}). Since the separatrices are themselves field lines, two x-points joined by a separatrix must have the same value of $A$.

In a 2D field, changes in topology---i.e., changes in the amount of flux in each region of the partition---can take place only at null points \citep{hornig2007}. Traditionally, reconnection counts only changes in $A$ at x-points, indicating the transfer of flux between distinct regions, and not at o-points, where changes in $A$ represent only the creation/annihilation of flux within a single region. In this way one can define a global partition reconnection rate
\begin{equation}
\Delta\Phi_{\rm P} = \sum_{{\bf h}_i}\left|\frac{dA({\bf h}_i)}{dt}\right|,
\end{equation}
where the sum is over all x-points ${\bf h}_i$. Notice that we can measure the reconnection rate completely knowing only the values of $A$ at the null points, with no need to know either the partition fluxes or the structure of the field (except for the spatial derivatives at each null point, in order to determine the hyperbolicity).

%%%%%%%%%%%%%%%%%%%%
\section{Generalized Flux Function} \label{sec:gen}

In a general 3D field, we can no longer write ${\bf B}$ in terms of a 2D function. But this does not prevent us from constructing a 2D function to measure magnetic flux. In this section, we will show how to construct a generalized flux function ${\cal A}(x,y)$ that retains a number of the properties of the 2D flux function $A(x,y)$. Our domain is a bounded region in $\mathbb{R}^3$ between $z=0$ and $z=1$, with all field lines connecting from the lower to the upper boundary.

As a simple generalization of $A(x,y)$, we might consider a function $f(x,y)=\int_0^1{\bf A}\cdot{\bf e}_z\,dz$, i.e. the integral along vertical lines of the vector potential ${\bf A}$ (where ${\bf B}=\nabla\times{\bf A})$. The difference in $f(x,y)$ between two points $(x_1,y_1)$ and $(x_2,y_2)$ would then give the flux through a vertical surface, analogous to the 2D case. However, this function $f(x,y)$ does not retain the ideal invariant property of $A(x,y)$, which is vital to define any meaningful reconnection rate.

To construct a flux function that \emph{is} an ideal invariant, we make a simple modification and integrate ${\bf A}$ along magnetic field lines rather than vertical lines. For a point $(x,y)$ on the lower boundary, denote the field line starting at $(x,y)$ by ${\bf F}_z(x,y)$. In other words,
\begin{equation}
\frac{\partial {\bf F}_z(x,y)}{\partial z} = \frac{{\bf B}({\bf F}_z(x,y))}{B_z({\bf F}_z(x,y))}, \quad {\rm with}\quad  {\bf F}_0(x,y) = (x,y).
\end{equation}
The subscript $z$ indicates that we have chosen to parametrise the field line by the vertical coordinate $z$. With this notation, we may define the \emph{generalized flux function} as a function on the lower boundary $z=0$:
\begin{equation}
{\cal A}(x,y) = \int_0^1{\bf A}\big({\bf F}_z(x,y)\big)\cdot\frac{{\bf B}\big({\bf F}_z(x,y)\big)}{B_z\big({\bf F}_z(x,y)\big)}\,dz.
\label{eqn:curlya}
\end{equation}

We consider in this paper only periodic fields where $B_z(x,y,1)=B_z(x,y,0)$, and we impose the gauge condition that ${\bf A}\times{\bf e}_z$ is periodic. We are still free to impose a gauge transformation ${\bf A}\rightarrow{\bf A}'+\nabla\chi$ providing that $\chi(x,y,1)=\chi(x,y,0) + \chi_0$ with $\chi_0$ constant. We impose the further gauge condition $\chi_0=0$, leaving the function $\chi(x,y,z)$ free for $0\leq z<1$.

\begin{figure}[htbp]
\begin{center}
\includegraphics[width=\columnwidth]{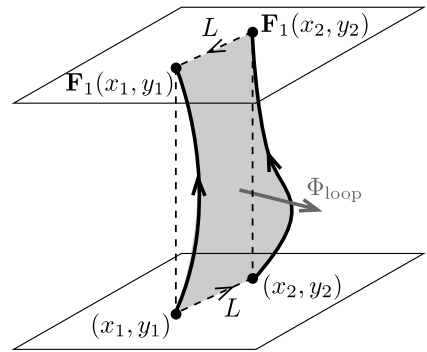}
\caption{The flux $\Phi_{\rm loop}$ through the surface defined by two fixed points $(x_1,y_1)$, $(x_2,y_2)$ is measured by ${\cal A}(x_2,y_2)-{\cal A}(x_1,y_1)$, because the integrals of ${\bf A}$ along line $L$ on $z=0$ and $z=1$ are equal and opposite.}
\label{fig:loop}
\end{center}
\end{figure}

Under a gauge transformation, the function ${\cal A}(x,y)$ becomes
\begin{equation}
{\cal A}'(x,y) = {\cal A}(x,y) + \chi\big({\bf F}_1(x,y)\big) - \chi{(x,y,0)},
\label{eqn:curlyatrans}
\end{equation}
so it is not gauge invariant in general. But at fixed points, where ${\bf F}_1(x,y)=(x,y)$, the last two terms in \eqref{eqn:curlyatrans} cancel and ${\cal A}(x,y)$  becomes gauge invariant. Thus differences in ${\cal A}$ between fixed points are well-defined, and correspond to physical fluxes (Figure \ref{fig:loop}), in analogy to the 2D case. We argue in this paper that these physical fluxes defined by values of ${\cal A}$ at fixed points form a natural partition of the 3D magnetic field. In fact, the values of ${\cal A}$ at non-fixed points may also be given physical meaning if one fixes the gauge in a particular way; this is beyond the scope of this paper and will be addressed in future.

%%%%%%%%%%%%%%%%%%%%
%%%%%%%%%%%%%%%%%%%%
\section{General flux partition} \label{sec:partition}

We propose a simple generalization of the 2D case: partition the flux in a 3D field by the hyperbolic fixed points of the field line mapping ${\bf F}_1(x,y)$ and their global manifolds. The physical nature of the partition is explored in this section, while the partition reconnection rate is defined in Section \ref{sec:reco}.

A 2D mapping may be viewed as a discrete-time dynamical system, and we will use mathematical methods developed for such systems. For more details see \citet{guckenheimer1983} or \citet{wiggins1992}.

Analogous to an x-point in a 2D vector field, a fixed point ${\bf x}_0$ of a 2D mapping is said to be \emph{hyperbolic} when the eigenvalues $\lambda_s$, $\lambda_u$ of the Jacobian matrix $J_{ij}=\partial F_{1,i}/\partial F_{1,j}$ at ${\bf x}_0$ satisfy $|\lambda_s|<1<|\lambda_u|$. As with the x-point, the associated eigenvectors define linear subspaces, and the map ${\bf F}_1$ has global stable and unstable manifolds $W^s({\bf x}_0)$, $W^u({\bf x}_0)$ that are tangent to these linear subspaces at ${\bf x}_0$. There are two branches of each manifold for each hyperbolic fixed point.

By definition, $W^s({\bf x}_0)$ and $W^u({\bf x}_0)$ are invariant subspaces, meaning that if ${\bf x}\in W^s({\bf x}_0)$ then ${\bf F}_1({\bf x})\in W^s({\bf x}_0)$, and similarly for $W^u({\bf x}_0)$. Under the mapping ${\bf F}_1$, points on $W^s({\bf x}_0)$ move closer to ${\bf x}_0$ (along the curve), while those on $W^u({\bf x}_0)$ move further away. In the case of our magnetic field, $W^s({\bf x}_0)$ and $W^u({\bf x}_0)$ correspond to curves on the boundary $z=0$ (or equivalently $z=1$). Their invariance means that field lines starting on either manifold for $z=0$ must end on the same manifold for $z=1$. The union of such field lines therefore defines a magnetic surface in the 3D domain generated by each manifold.

%%%%%%%%%%%%%%%%%%%%%
\subsection{Integrable fields}

\begin{figure}[htbp]
\begin{center}
\includegraphics[width=\columnwidth]{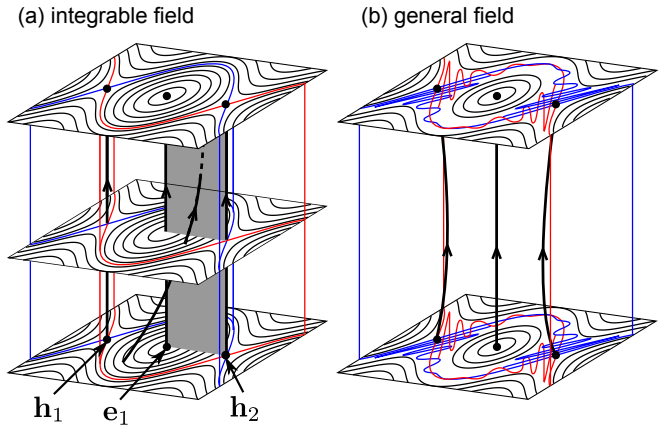}
\caption{Sketch of (a) an integrable field where field lines lie on flux surfaces, and (b) a more general 3D field created by perturbing the integrable field. Thick black lines show magnetic field lines, including the three fixed points of the integrable field (${\bf h}_1$, ${\bf h}_2$ hyperbolic and ${\bf e}_1$ elliptic), which persist in the general field. Red and blue curves show where the global manifolds intersect the boundary.}
\label{fig:islands}
\end{center}
\end{figure}

The simplest type of 3D field to understand is an \emph{integrable} field, where the field lines lie on a foliation of flux surfaces. Figure \ref{fig:islands}(a) shows an example of such a field defined by adding a uniform $z$-component to a 2D magnetic field. Field lines of the 3D field lie on vertical surfaces that project on to field lines of the 2D field. The three null points of the 2D field now correspond to vertical field lines, thus to fixed points (${\bf e}_1$, ${\bf h}_1$, ${\bf h}_2$). The separatrices of the 2D field correspond to the global manifolds of the 3D field line mapping.

What is the flux of the ``island'' containing ${\bf e}_1$? There are two natural fluxes: (1) the vertical flux through the lower boundary within this region, and (2) the horizontal flux crossing the grey-shaded vertical surface between the ${\bf e}_1$ and ${\bf h}_2$ field lines. The first flux does not exist in the original 2D field, but is measured straightforwardly from ${\bf B}\cdot{\bf n}$ on the lower boundary. The second flux is measured using the generalized flux function by the difference ${\cal A}({\bf e}_1) - {\cal A}({\bf h}_2)$. This is clearly analogous to the 2D case (Section \ref{sec:2d}). Note that, in this integrable field, ${\cal A}({\bf h}_1)={\cal A}({\bf h}_2)$, so the identical flux for this island region would be measured by ${\cal A}({\bf e}_1) - {\cal A}({\bf h}_1)$. The ``barrier'' around the island comprises two magnetic surfaces: (1) a branch of $W^u({\bf h}_1)$, which coincides with a branch of $W^s({\bf h}_2)$ (in red), and (2) a branch of $W^s({\bf h}_1)$, which coincides with a branch of $W^u({\bf h}_2)$ (in blue).

%%%%%%%%%%%%%%%%%
\subsection{Heteroclinic tangles}

\begin{figure}[htbp]
\begin{center}
\includegraphics[width=\columnwidth]{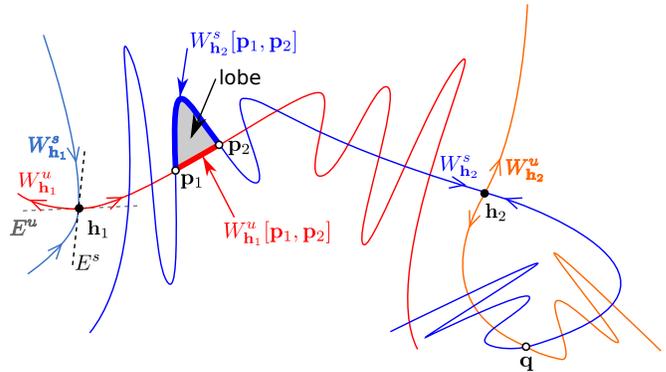}
\caption{Example and notation for a heteroclinic tangle between two hyperbolic fixed points ${\bf h}_1$ and ${\bf h}_2$. The manifolds are curtailed to finite length for clarity. Pip ${\bf q}$ is a homoclinic intersection. Pips ${\bf p}_1$, ${\bf p}_2$ are heteroclinic intersections, defining the lobe shaded in grey.}
\label{fig:hyperbolic}
\end{center}
\end{figure}

Unfortunately, the simplicity of the integrable case belies the complexity typical of a general 3D magnetic field. If a small $z$-dependent perturbation is applied to Figure \ref{fig:islands}(a), the three fixed points will persist and maintain their elliptic/hyperbolic character, but the regular global manifolds will break down into \emph{heteroclinic tangles} (Figure \ref{fig:islands}(b)). In this generic situation, the stable and unstable manifolds intersect transversally at discrete points, rather than coinciding to form regular separatrices as in the 2D or integrable cases. It follows from the uniqueness of field lines that an intersection can take place only between a stable manifold and an unstable manifold. Two stable manifolds can never intersect, nor can two unstable manifolds. An intersection between a stable manifold and an unstable manifold of the same fixed point is called a \emph{homoclinic point}, while an intersection between manifolds from different fixed points is a \emph{heteroclinic point} (Figure \ref{fig:hyperbolic}). In this paper, we shall not need to distinguish between the two, and will refer to both as heteroclinic points.

The key result about heteroclinic intersections, first recognised by Poincar\'{e}, is that a single intersection between two manifolds $W^s_{{\bf h}_1}$ and $W^u_{{\bf h}_2}$ implies the existence of an infinite number of intersections between these same two curves. This simply follows from the fact that the intersection point lies on both manifolds. It cannot be a fixed point, and every iterate must also lie on both manifolds, by definition. The infinite number of intersections as one approaches either of the fixed points ${\bf h}_1$, ${\bf h}_2$ leads to a very convoluted path of the manifold curves. Called a \emph{homoclinic tangle}, this is a major route to chaos in 2D mappings. This possibility of chaos in the field line mapping is the major factor that complicates the partitioning of flux in a 3D field.

%%%%%%%%%%%%%%%%%
\subsection{Partial barriers}

Since the global manifolds for a 3D field can be infinitely long (unlike in 2D), the regions of the flux partition must be defined by \emph{partial barriers}: curves comprising segments of one or more global manifolds, ending at hyperbolic fixed points \citep{beigie1994}.

To formally define a partial barrier, let $W^s_{{\bf h}}[{\bf x}_1,{\bf x}_2]$ denote the segment of $W^s_{{\bf h}}$ between two points ${\bf x}_1$, ${\bf x}_2$. Consider an intersection point ${\bf p}\in W^u_{{\bf h}_1}\cap W^s_{{\bf h}_2}$. This point ${\bf p}$ is a \emph{primary intersection point} or \emph{pip} if the segments $W^u_{{\bf h}_1}[{\bf h}_1,{\bf p}]$ and $W^s_{{\bf h}_2}[{\bf p},{\bf h}_2]$ intersect only at ${\bf p}$ (and possibly at ${\bf h}_1$ if ${\bf h}_2={\bf h}_1$; \citet{romkedar1990}). A partial barrier starts and ends at hyperbolic fixed points (possibly the same), and comprises one or more global manifold segments intersecting at pips. It includes no further fixed points.

\begin{figure}[htbp]
\begin{center}
\includegraphics[width=0.8\columnwidth]{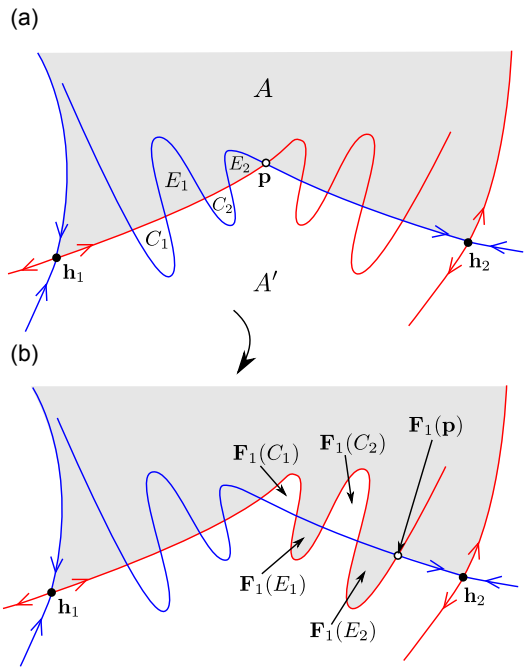}
\caption{A partial barrier and the turnstile lobes.}
\label{fig:transport}
\end{center}
\end{figure}

Figure \ref{fig:transport}(a) shows a partial barrier between hyperbolic fixed points ${\bf h}_1$ and ${\bf h}_2$, with two segments $W^u_{{\bf h}_1}[{\bf h}_1,{\bf p}]$ and $W^s_{{\bf h}_2}[{\bf p},{\bf h}_2]$ intersecting at pip ${\bf p}$. The barrier separates the shaded region $A$ from the unshaded region $A'$. There is nothing special about this choice of pip: choosing a different pip would redefine the barrier and also the shape of regions $A$ and $A'$. But the partition fluxes are defined only by ${\cal A}$ at the fixed points, so are independent of the choice of partial barrier.

The barrier in Figure \ref{fig:transport} is called ``partial'' because certain field lines cross it in the mapping ${\bf F}_1$. While no magnetic field line may cross the magnetic surface generated by each global manifold, this does not prevent field lines from crossing the partial barrier if it is made up of more than one global manifold. In the remainder of this section we show that the flux crossing the barrier in each direction under ${\bf F}_1$ is well-defined (independent of the choice of pip ${\bf p}$), and further that the net flux crossing the barrier is simply ${\cal A}({\bf h}_2)-{\cal A}({\bf h}_1)$.

The key to understanding which field lines cross a partial barrier is \emph{lobe dynamics}\citep{mackay1984,romkedar1990,meiss1992,beigie1994}. A \emph{lobe} is a closed region bounded by the segments $W^u_{{\bf h}_1}[{\bf p}_1,{\bf p}_2]$, $W^s_{{\bf h}_2}[{\bf p}_1,{\bf p}_2]$ between two adjacent pips ${\bf p}_1$, ${\bf p}_2$ (e.g., Figure \ref{fig:hyperbolic}). The important dynamical rules governing lobes are\citep{romkedar1990}:
\begin{enumerate}
\item Lobes map to lobes under ${\bf F}_1$. This follows from continuity of the mapping and the fact that $W^u$ and $W^s$ are invariant manifolds that field lines cannot cross.
\item Ordering of points on $W^u$ and $W^s$ is maintained, so for a given pair of intersecting manifolds, there are a fixed number $m$ of lobes lying between ${\bf p}$ and ${\bf F}_1({\bf p})$, the same for any pip ${\bf p}$.
\end{enumerate}
In our case, ${\bf F}_1$ is orientation-preserving ($|J|>0$ because $B_z>0$), so $m$ must be even.

Consider again Figure \ref{fig:transport}, where $m=4$. In the mapping ${\bf F}_1$, the two lobes $E_1$, $E_2$ cross from $A$ to $A'$, while the two lobes $C_1$, $C_2$ cross from $A'$ to $A$. These four lobes, which are precisely those lying between ${\bf p}$ and ${\bf F}_1({\bf p})$, are the \emph{turnstile lobes}: they contain exactly those points which cross the partial barrier under ${\bf F}_1$. The flux in a lobe $L$ is measured by integrating $\Phi(L) = \int_L B_z(x,y,0)\,dxdy$ on $z=0$. What happens if we choose a different pip to define the partial barrier? The turnstile would then comprise different lobes. But the dynamical rules above guarantee that \emph{there would still be four turnstile lobes, and their fluxes would be the same as for the original choice of pip}.

\begin{thm}[Net flux]
Let ${\bf p}$ be a pip of $W^u_{{\bf h}_1}$ and $W^u_{{\bf h}_2}$ defining a partial barrier between regions $A$ and $A'$ (oriented as in Figure \ref{fig:transport}). For $i=1,..,m/2$, let $E_i$ be the lobes mapped from $A$ to $A'$ by ${\bf F}_1$, and $C_i$ the lobes mapped from $A'$ to $A$. Then
\[
{\cal A}({\bf h}_2) - {\cal A}({\bf h}_1) = \sum_{i=1}^{m/2}\big(\Phi(E_i) - \Phi(C_i)\big),
\]
and this sum is independent of the choice of pip ${\bf p}$.
\label{thm:netflux}
\end{thm}

\begin{figure}[htbp]
\begin{center}
\includegraphics[width=\columnwidth]{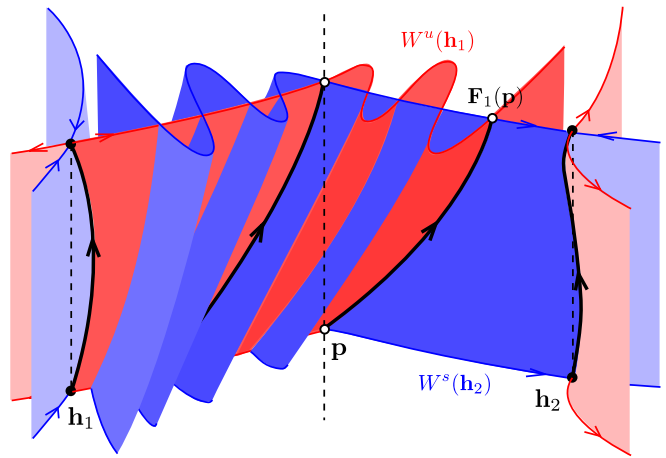}
\caption{Sketch of the magnetic surfaces in the 3D domain generated by field lines from $W^u_{{\bf h}_1}$ (in red) and $W^s_{{\bf h}_2}$ (in blue). Some important magnetic field lines are shown in black.}
\label{fig:3dcartoon}
\end{center}
\end{figure}

\begin{proof}

We already know from the rules of lobe dynamics that the sum is independent of the choice of pip.

The sketch in Figure \ref{fig:3dcartoon} illustrates the magnetic surfaces generated by $W^u_{{\bf h}_1}$ and $W^u_{{\bf h}_2}$, for the barrier in Figure \ref{fig:transport}. To derive our result, we consider two closed loops, one lying on each of these surfaces. Start with the following closed loop on the surface generated by $W^u_{{\bf h}_2}$:
\begin{equation}
L_u \equiv {\bf F}_z({\bf p})\cup W^u_{{\bf h}_1}[{\bf F}_1({\bf p}),{\bf h}_1]\cup {\bf F}^{-1}_z({\bf h}_1)\cup W^u_{{\bf h}_1}[{\bf h}_1,{\bf p}].
\end{equation}
Here the notation ${\bf F}_z({\bf p})$ means the field line traced from ${\bf p}$ on the lower boundary to ${\bf F}_1({\bf p})$ on the upper boundary, and ${\bf F}_z^{-1}$ means a field line traced downward from the upper boundary to the lower boundary.

Now form a closed loop on the $W^s_{{\bf h}_1}$ surface:
\begin{equation}
L_s \equiv {\bf F}_z({\bf p})\cup W^s_{{\bf h}_2}[{\bf F}_1({\bf p}),{\bf h}_2]\cup {\bf F}^{-1}_z({\bf h}_2)\cup W^s_{{\bf h}_1}[{\bf h}_2,{\bf p}].
\end{equation}
The integral of ${\bf A}$ around each loop must vanish, so, using periodicity of ${\bf A}\times{\bf e}_z$,
\begin{align}
{\cal A}({\bf p}) - \int_{W^u_{{\bf h}_1}[{\bf p},{\bf F}_1({\bf p})]}{\bf A}\cdot\,d{\bf l} - {\cal A}({\bf h}_1) &= 0,\label{eqn:loop1}\\
{\cal A}({\bf p}) - \int_{W^s_{{\bf h}_2}[{\bf p},{\bf F}_1({\bf p})]}{\bf A}\cdot\,d{\bf l} - {\cal A}({\bf h}_2) &= 0.\label{eqn:loop2}
\end{align}
Subtracting \eqref{eqn:loop2} from \eqref{eqn:loop1} yields
\begin{equation}
{\cal A}({\bf h}_2) - {\cal A}({\bf h}_1) = \int_{W^u_{{\bf h}_1}[{\bf p},{\bf F}_1({\bf p})]}{\bf A}\cdot\,d{\bf l} - \int_{W^s_{{\bf h}_2}[{\bf p},{\bf F}_1({\bf p})]}{\bf A}\cdot\,d{\bf l}.
\label{eqn:areas}
\end{equation}
The right-hand side is the magnetic flux through a closed loop in the plane $z=0$ (or $z=1$) encircling all of the turnstile lobes. Since the ${\bf F}_1(E_i)$ are encircled anticlockwise and the ${\bf F}_1(C_i)$ are encircled clockwise, the result follows from Stokes' theorem.
\end{proof}

Theorem \ref{thm:netflux} shows that, for two hyperbolic points connected by a partial barrier, the difference in their values of ${\cal A}$ is precisely the net flux crossing this partial barrier. If ${\bf h}_2={\bf h}_1$, i.e. the two manifolds belong to the \emph{same} fixed point, then there can be no net flux across the barrier. In the limiting case that ${\bf h}_1$ and ${\bf h}_2$ are connected by a regular separatrix (i.e., the two global manifolds coincide exactly, as in a 2D field), there are effectively infinitely many lobes of zero area. In this limiting case, Theorem \ref{thm:netflux} reduces to ${\cal A}({\bf h}_2)={\cal A}({\bf h}_1)$, as in 2D.

For simplicity, our illustrations avoid secondary intersections between lobes (see \citet{romkedar1990}). However, even in the presence of secondary intersections, one may show that ${\cal A}({\bf h}_2) - {\cal A}({\bf h}_1)$ gives the net flux across the partial barrier.

%%%%%%%%%%%%%%%%%
\subsection{A region of the general flux partition}

\begin{figure}[htbp]
\begin{center}
\includegraphics[width=\columnwidth]{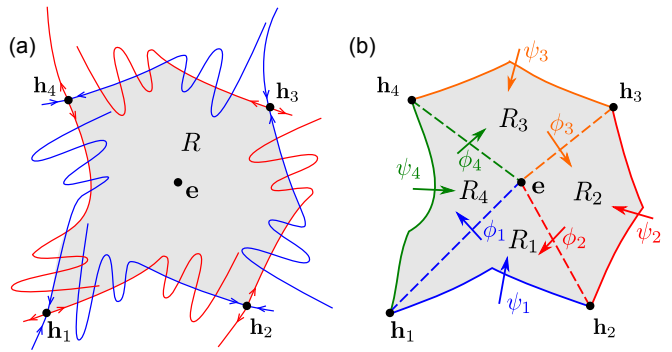}
\caption{A region $R$ of the general flux partition (a), and decomposition into subregions $R_i$ (b).}
\label{fig:regions}
\end{center}
\end{figure}

To get a feeling for the nature of our new flux partition, let us examine the region $R$ in Figure \ref{fig:regions}(a), whose boundary is a chain of partial barriers comprising alternating segments of $W^u$ and $W^s$ from four hyperbolic fixed points ${\bf h}_i$, $i=1,\ldots, 4$. The exact definition of each partial barrier is non-unique, owing to the freedom of choice of defining pip, but this does not affect any of the fluxes that we will describe here.

Since $R$ is simply connected it must contain an elliptic fixed point ${\bf e}$, because the topological degree of ${\bf F}_1$ on $R$ is 1, by definition of the boundary. In this respect, the situation is analogous to a similar region in a 2D magnetic field, which must contain an o-point (Figure \ref{fig:2d}). However, there is a key difference when we try to define a ``flux'' of the region $R$. In the 2D field, the partial barriers would be replaced by regular separatrices, and the flux function $A$ would have the same value at each ${\bf h}_i$. The flux of the region would then be unambiguously defined as $A({\bf e})-A({\bf h}_1)$. In the 3D case, such a unique flux cannot be defined.

To see this, consider the differences in ${\cal A}$ between each pair of fixed points, which define 8 physical fluxes, shown in Figure \ref{fig:regions}(b):
\begin{align*}
\psi_1 &= {\cal A}({\bf h}_2)-{\cal A}({\bf h}_1), & \qquad \phi_1 &= {\cal A}({\bf e}) - {\cal A}({\bf h}_1),\\
\psi_2 &= {\cal A}({\bf h}_3)-{\cal A}({\bf h}_2), & \qquad \phi_2 &= {\cal A}({\bf e}) - {\cal A}({\bf h}_2),\\
\psi_3 &= {\cal A}({\bf h}_4)-{\cal A}({\bf h}_3), & \qquad \phi_3 &= {\cal A}({\bf e}) - {\cal A}({\bf h}_3),\\
\psi_4 &= {\cal A}({\bf h}_1)-{\cal A}({\bf h}_4), & \qquad \phi_4 &= {\cal A}({\bf e}) - {\cal A}({\bf h}_4).
\end{align*}
The $\psi_i$ are the net fluxes through each partial barrier, while the $\phi_i$ measure fluxes across surfaces in the domain. But these 8 fluxes are not all independent. Dividing $R$ into 4 subregions $R_i$, as in Figure \ref{fig:regions}(b), the net flux into each $R_i$ must vanish (for a periodic field), so we have the constraints
\begin{equation}
\label{constraints}
\psi_i = \phi_i-\phi_{i+1}, \qquad \textrm{for $i=1,\ldots,4$}.
\end{equation}
Summing all of these equations leads to $\sum_i\psi_i=0$, expressing conservation of flux in the full region $R$. It is clear from (\ref{constraints}) that differences between the $\phi_i$ relate to net flux through the partial barriers. If all $\psi_i=0$, as in a 2D field, all of the $\phi_i$ must be equal, giving us our uniquely defined flux. But if any of the $\psi_i$ are non-zero, there is no meaningful single flux in $R$. 

Interestingly, we see that a change in ${\cal A}({\bf e})$ adds the same amount to each $\phi_i$. So although the structure is not simple enough to define a unique flux in $R$, there is a unique \emph{reconnected} flux. This emphasizes that it is the values of ${\cal A}$ at fixed points that define our flux partition, not the individual fluxes $\psi_i$, $\phi_i$, which are not independent. Note that this change in ${\cal A}({\bf e})$ does not affect any of the $\psi_i$, so it represents a purely \emph{local} non-ideal event within the region $R$. By contrast, changes in ${\cal A}({\bf h}_i)$ affect more than one region of the partition, giving reconnection in the usual sense.

%%%%%%%%%%%%%%%%%%%%
%%%%%%%%%%%%%%%%%%%%
\section{Measuring reconnection} \label{sec:reco}

Measuring the partition reconnection rate for our general flux partition is straightforward using the generalized flux function ${\cal A}$. As in the 2D case, the partition fluxes are defined entirely by the values of ${\cal A}$ at fixed points. In a general 3D field line mapping with chaotic regions, we cannot uniquely define the regions of the partition, because the definition of partial barriers is non-unique. But to measure reconnection we require not a partition of space into regions but a partition of \emph{flux}. This we have shown to be well-defined.

In the same way as the 2D case, we can define a global partition reconnection rate by summing over hyperbolic fixed points ${\bf h}_i$:
\begin{equation}
\Delta\Phi_{\rm P} = \sum_{{\bf h}_i}\left|\frac{d{\cal A}({\bf h}_i)}{dt}\right|.
\end{equation}

How does this partition reconnection rate relate to the definition of reconnection using integrated $E_{||}$? In the latter definition, reconnection can occur anywhere with non-zero $E_{||}$: a global reconnection rate is defined\citep{schindler1988} by identifying distinct reconnection regions as local maxima of $E_{||}$. These sites need not coincide with fixed point field lines, so the global reconnection rates from the two methods may differ. The example in Section \ref{sec:example} will demonstrate this. In fact, the change in ${\cal A}$ at a fixed point corresponds to the integral of $E_{||} $ along the fixed point field line itself, as we now show. 

Let ${\bf x}_0$ be a fixed point of the field line mapping ${\bf F}_1$, either hyperbolic or elliptic. We consider a general non-ideal evolution with an Ohm's law of the form
\begin{equation}
{\bf E} + {\bf v}\times{\bf B} = {\bf R},
\label{eqn:ohm}
\end{equation}
where ${\bf R}$ is any non-ideal term (for example, in resistive-MHD, ${\bf R}={\bf j}/\sigma$). In our magnetic field, where there are no closed field lines, we may write ${\bf R} = \nabla\psi + {\bf u}\times{\bf B}$, where
\begin{equation}
\psi\big({\bf F}_z(x,y)\big) = \int_{(x,y)}^{{\bf F}_z(x,y)}{\bf R}\cdot\,d{\bf l}
\end{equation}
is integrated along a field line, and ${\bf u} = {\bf B}\times({\bf R}-\nabla\psi)/B^2$. 
Letting ${\bf w}={\bf v}-{\bf u}$, we may write Ohm's law \eqref{eqn:ohm} as
\begin{equation}
{\bf E} + {\bf w}\times{\bf B} = \nabla\psi,
\end{equation}
indicating that the field lines are frozen-in with a velocity ${\bf w}$, which differs in general\citep{hornig2007} from the plasma velocity ${\bf v}$.

Faraday's law then implies that
\begin{equation}
\frac{\partial {\bf A}}{\partial t} - {\bf w}\times\nabla\times{\bf A} = -\nabla(\phi + \psi),
\label{eqn:dadt}
\end{equation}
where $\phi({\bf x},t)$ is the electrostatic potential. Following a field line at velocity ${\bf w}$, the rate of change of ${\cal A}$ is
\begin{align}
\frac{\partial {\cal A}}{\partial t} &+ {\bf w}\cdot\nabla{\cal A} = \frac{d}{dt}\int{\bf A}\cdot\,d{\bf l}\\
&= \int\left(\frac{\partial {\bf A}}{\partial t} - {\bf w}\times\nabla\times{\bf A} + \nabla({\bf w}\cdot{\bf A}) \right)\cdot\,d{\bf l}\\
&= \big({\bf w}\cdot{\bf A} - \phi - \psi \big)\big|_{(x,y)}^{{\bf F}_1(x,y)},
\end{align}
using \eqref{eqn:dadt}. At a fixed point, ${\bf F}_1({\bf x}_0)={\bf x}_0$, and we now choose the gauge $\phi$ so that $\phi= {\bf w} \cdot {\bf A}$. This ensures that ${\cal A}$ becomes an ideal invariant whenever ${\bf R}=0$, and leads to
\begin{equation}
\frac{\partial {\cal A}}{\partial t} + {\bf w}\cdot\nabla{\cal A}  = -\int_{{\bf x}_0}^{{\bf F}_1({\bf x}_0)}{\bf E}\cdot\,d{\bf l}.
\end{equation}
In other words, the rate of change of ${\cal A}$ following a fixed point corresponds to the integrated parallel electric field along the fixed point field line.

%%%%%%%%%%%%%%%%%%%%%%%%%
%%%%%%%%%%%%%%%%%%%%%%%%%
\section{Example}  \label{sec:example}

To illustrate the ideas developed above, we present a particular example of a 3D magnetic field. The basic field (Section \ref{sec:basic}) is given by an analytical expression, and is chosen to be generic in that the field line mapping contains both regular and chaotic regions, and the global manifolds form heteroclinic tangles. In Section \ref{sec:added} we perform a simple experiment where a growing toroidal flux ring is added to the basic field. This demonstrates how our flux partition responds to localized 3D reconnection.

\subsection{Structure of the basic field} \label{sec:basic}

\begin{figure*}[htbp]
\begin{center}
\includegraphics[width=\textwidth]{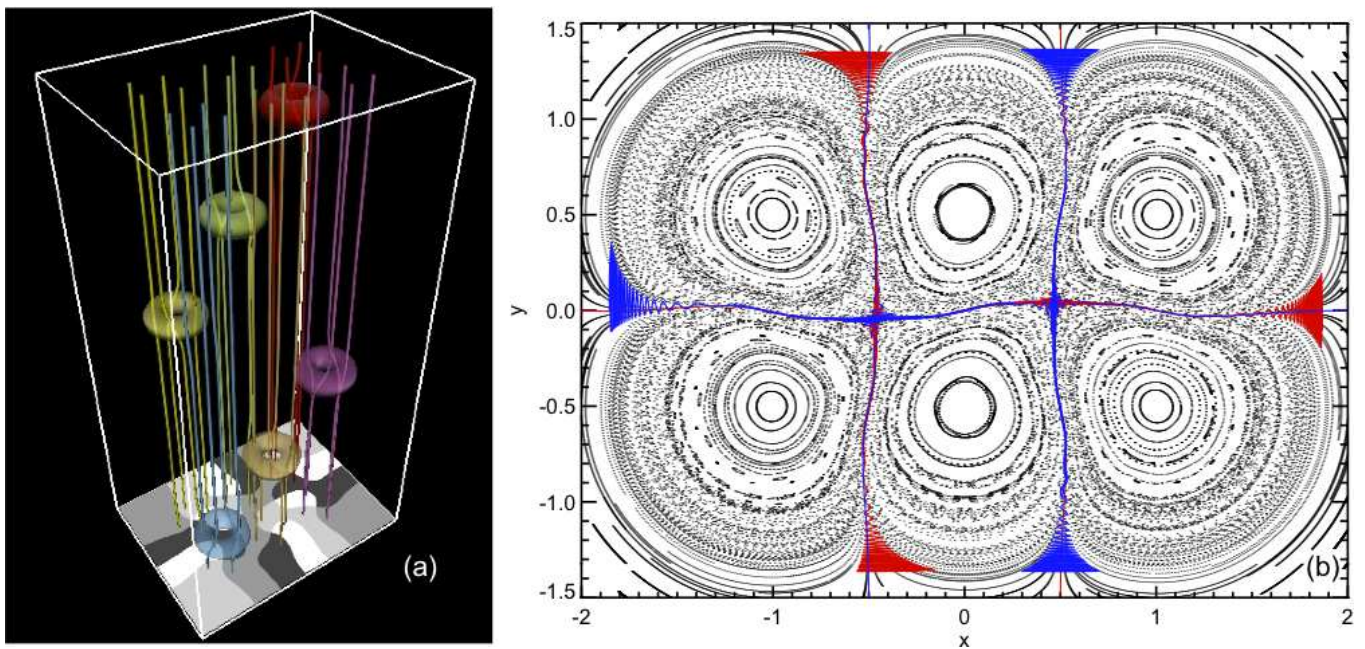}
\caption{The 6-roll magnetic field. The \emph{left} panel shows a 3D visualization (isosurfaces at $B_x^2 + B_y^2=0.003$ identify the toroidal flux rings), while the \emph{right} panel shows a Poincar\'{e} return map with the calculated global stable/unstable manifolds shown in blue/red respectively. The greyscale image on the lower boundary of the 3D visualization shows the ``color map'' used to identify fixed points.}
\label{fig:sixroll_field}
\end{center}
\end{figure*}

The basic field comprises six isolated magnetic flux rings, superimposed on a uniform vertical field $B_z=1$. The $i$th flux ring is derived from a vector potential
\begin{equation}
{\bf A}_i=a_ik_i\exp\left(-\frac{(x-x_i)^2 + (y-y_i)^2}{a_i^2} - \frac{(z-z_i)^2}{l_i^2} \right){\bf e}_z,
\end{equation}
where the centre of the ring is at $(x_i,y_i,z_i)$, the parameter $k_i$ controls the flux, and the parameters $a_i$ and $l_i$ control the radial and vertical extents respectively \citep{wilmotsmith2009}. Thus
\begin{equation}
{\bf B} = \nabla\times\left( -\frac{y}{2}{\bf e}_x + \frac{x}{2}{\bf e}_y + \sum_{i=1}^6 {\bf A}_i \right),
\label{eqn:basic}
\end{equation}
where we choose the parameter sets 
\begin{eqnarray*}
(x_i,i=1,\dotsc, 6)&=&(1,0,-1,1,0,-1),\\
(y_i,i=1,\dotsc, 6)&=&(0.5,0.5,0.5,-0.5,-0.5,-0.5),\\
(z_i,i=1,\dotsc, 6)&=&(-20,-12,4,4,12,20),\\
(k_i,i=1,\dotsc, 6)&=&(1,-1,1.1,-1,1,-1)k_0
\end{eqnarray*}
with $k_0=0.08$, all $a_i=0.3\sqrt{2}$ and $l_i=2$. Notice that one of the rings has larger $|k_i|$: this creates an asymmetry in the flux partition leading to a net flux across certain partial barriers.

Figure \ref{fig:sixroll_field}(a) illustrates this magnetic field. Since both ${\bf A}$ and ${\bf B}$ are periodic, we can iterate the field line mapping ${\bf F}_1(x,y)$ to produce a Poincar\'{e} plot (Figure \ref{fig:sixroll_field}(b)). This reveals that the field is structured into six regular elliptic regions---corresponding to the $(x,y)$ locations of the six flux rings---separated by bands of chaotic field lines. 

The 2D greyscale image on the base of Figure \ref{fig:sixroll_field}(a) shows a ``color map''\citep{polymilis2003} of the direction of ${\bf F}_1(x,y)-(x,y)$. We used this to identify fixed points as locations where all four colors meet, and to identify the Poincar\'{e} index of each fixed point from the surrounding color sequence\citep{yeates2010,yeates2011}. The precise locations (Table \ref{tab:fp}) were found by Newton-Raphson iteration using the color map as a first guess. There are six elliptic fixed points, at the centre of each regular region, and two hyperbolic fixed points between. Table \ref{tab:fp} also shows the value of the generalized flux function ${\cal A}(x,y)$ at each fixed point, calculated by numerically integrating ${\bf A}$ along the appropriate magnetic field line from $z=-24$ to $z=24$.

\begin{table}[htdp]
\caption{Fixed points in the basic field \eqref{eqn:basic}.}
\begin{center}
\begin{tabular}{ccrrr}
Point & Poincar\'{e} index & $x$ & $y$ & ${\cal A}(x,y)$\\
\hline
${\bf e}_1$ & 1 & $1.0058$ & $0.5007$ & $0.1191$\\
${\bf e}_2$ & 1 & $-0.0021$ & $0.5083$ & $-0.1189$\\
${\bf e}_3$ & 1 & $-1.0003$ & $0.5003$ & $0.1314$\\
${\bf e}_4$ & 1 & $-1.0061$ & $-0.5010$& $-0.1191$\\
${\bf e}_5$ & 1 & $0.0024$ & $-0.5081$ & $0.1190$\\
${\bf e}_6$ & 1 & $1.0007$ & $-0.5007$ & $-0.1194$\\
${\bf h}_1$ & -1 & $-0.4623$ & $-0.0457$ & $0.0007$\\
${\bf h}_2$ & -1 & $0.4663$ & $0.0410$ & $0$ 
\end{tabular}
\end{center}
\label{tab:fp}
\end{table}%

The red and blue curves in Figure \ref{fig:sixroll_field}(b) show the global manifolds of the fixed points ${\bf h}_1$ and ${\bf h}_2$. These have been ``grown'' numerically up to a finite length using the method of \citet{krauskopf1998} (see also \citet{england2004}). As expected in a generic 3D mapping, the manifolds do not describe regular separatrices, but have degenerated into heteroclinic tangles. In addition to the manifolds of ${\bf h}_1$ and ${\bf h}_2$, we also show the corresponding manifolds emanating from the six hyperbolic points at infinity (effectively on the boundaries of this plot).

\begin{figure*}[htbp]
\begin{center}
\includegraphics[width=\textwidth]{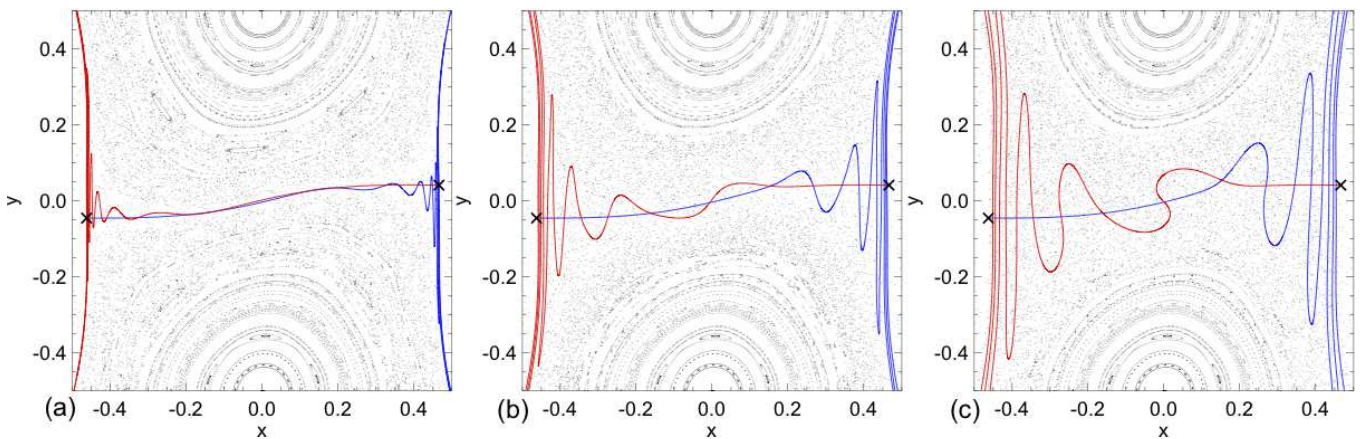}
\caption{Enlargement of the central partial barrier in Figure \ref{fig:sixroll_field}(b), showing the lobes in (a) the basic field, (b) Case 2 (Section \ref{sec:added}) at $t=2$, and (c) Case 3 at $t=6$.}
\label{fig:sixroll_reco}
\end{center}
\end{figure*}

An enlargement of the partial barrier between ${\bf h}_1$ and ${\bf h}_2$ is shown in Figure \ref{fig:sixroll_reco}(a). By identifying a pip ${\bf p}$ and computing ${\bf F}_1({\bf p})$, we find that $m=2$ for this example: i.e., one turnstile lobe crosses the partial barrier in each direction. It is apparent that the areas of these two lobes are unequal, i.e., there is a net flux across this partial barrier. Numerical integration shows that the two lobe fluxes are approximately $0.0008$ and $0.0001$, and indeed ${\cal A}({\bf h}_2)-{\cal A}({\bf h}_1) = 0.0007$, thus verifying Theorem \ref{thm:netflux} for this example. Repeating the calculation for the other partial barriers, we find that those connecting ${\bf h}_1$ with the boundary each have a net flux of $0.0007$, while those connecting ${\bf h}_2$ with the boundary each have zero net flux. This is consistent with the values of ${\cal A}$ for the two hyperbolic points (at infinity, ${\cal A}=0$). One can think of a net chaotic flux of $0.0007$ encircling the left-hand hyperbolic point, crossing all four of its attached partial barriers. (These barriers must all have the same net flux since ${\bf F}_1$ is area preserving.)

\subsection{Effect of an isolated reconnection region} \label{sec:added}

To illustrate several key properties of our general flux partition, we consider the effect of adding a gradually strengthening seventh flux ring to the basic field. This models the topological effect of a localised three-dimensional (non-null) diffusion region, as modelled by \citet{hornig2003} and studied in the framework of general magnetic reconnection \citep{hesse2005}.

Specifically, 
\begin{equation}
{\bf B} = \nabla\times\left( -\frac{y}{2}{\bf e}_x + \frac{x}{2}{\bf e}_y + \sum_{i=1}^7 {\bf A}_i \right),
\label{eqn:pert}
\end{equation}
where ${\bf A}_i$ are the same as the basic field for $i=1,\ldots,6$, and the new ring has parameters $z_7=28$, $a_7=0.1$, $l_7=1$, and $k_7=0.01t$. The dependence of $k_7$ on time $t$ causes a gradual increase in the ring's azimuthal magnetic flux from $\Phi_7=0$ at $t=0$ to $\Phi_7=\sqrt{\pi}a_7l_7k_7(t)$ at time $t$. We shall illustrate how the reconnection associated with this new flux ring affects the partition fluxes for two differing locations $(x_7,y_7)$.

\begin{figure*}[htbp]
\begin{center}
\includegraphics[width=\textwidth]{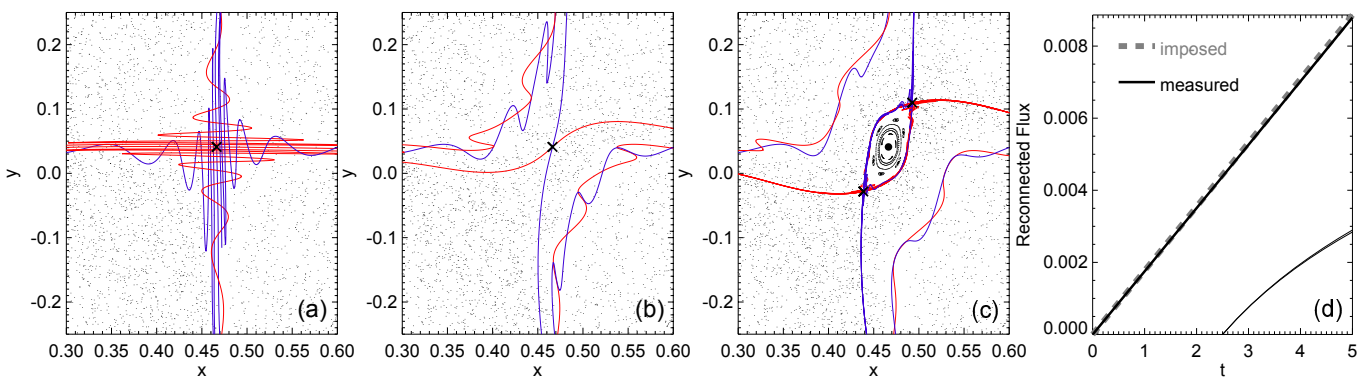}
\caption{Perturbation of the Poincar\'{e} plot and global manifolds in Case 1, at (a) $t=0$ (basic field), (b) $t=2$ and (c) $t=4$. Panel (d) compares the reconnected flux measured using ${\cal A}({\bf h}_2)$ (thick solid line) to the imposed flux $\Phi_7$ (grey dashed line): the two curves coincide. The thin lines show the ${\cal A}$ values at the two new hyperbolic points following the bifurcation at $t\approx 2.5$.}
\label{fig:sixroll_case4}
\end{center}
\end{figure*}

\emph{Case 1: At fixed point ${\bf h}_2$} ($x_7=0.4663 $, $y_7= 0.0410$).
The resulting perturbation is shown in Figure \ref{fig:sixroll_case4}. The fixed point remains at the same position, and for small $t$ remains hyperbolic, though the structure of the global manifolds underlying the chaotic region is altered. At $t\approx 2.5$, there is a pitchfork bifurcation: the original fixed point becomes elliptic and a pair of new hyperbolic fixed points are formed. The flux $\Phi_7$ of the new flux ring has become strong enough to perturb the field and create a new elliptic region.

Note that, throughout this evolution, the change in ${\cal A}({\bf h}_2)$ is exactly equal to the rate of increase of $\Phi_7$ (Figure \ref{fig:sixroll_case4}(d)). All of the new flux is counted by our partition reconnection rate, because the fixed point field line passes through the centre of the reconnection region. In general, a reconnection site will not be aligned with the fixed point field lines in this way, so the full imposed flux will not be measured by our partition reconnection rate. However, as suggested by the bifurcation in this example, if enough flux is reconnected then the structure of the underlying field will be modified, creating new fixed points that subsequently measure the new flux.

\emph{Case 2: On a partial barrier} ($x_7=y_7=0$).
Here the flux ring modifies field lines in the lobes of a partial barrier (Figure \ref{fig:sixroll_reco}). Figures \ref{fig:sixroll_reco}(b) and (c) show that the lobes grow as the ring flux increases. This implies that more flux crosses the partial barrier in each direction. But the \emph{net} flux across the barrier remains invariant because neither fixed point field line passes through the reconnection region, so that the fixed point locations and ${\cal A}$ values cannot change. Notice that, even though the mapping near ${\bf h}_1$ or ${\bf h}_2$ is unperturbed, the nearby lobes change significantly, because the global manifold passes through the reconnection region. Numerical computation of the areas of the two turnstile lobes reveals that $0.58(E_1 + C_1) \approx \Phi_7$, roughly corresponding to a simple picture of the new flux being counted twice in the lobes: once in each direction across the partial barrier. But this reconnection has not changed the partition fluxes as defined by ${\cal A}$ at fixed points, hence the partition reconnection rate is zero.

To summarise, these examples illustrate three important properties of an isolated reconnection region:
\begin{enumerate}
\item The position of a localised reconnection region within the background field determines its \emph{topological effectiveness}: i.e., the extent to which it changes the partition fluxes, as measured by ${\cal A}$ at fixed points. Indeed, the fluxes in the partition can change only if a fixed point field line passes through the reconnection region.
\item If the reconnected flux becomes large enough, the reconnection region may perturb the original field sufficiently that new fixed points whose field lines pass through the reconnection region are created. Thus it will become visible to the flux partition.
\item If field lines from the lobes of a heteroclinic tangle pass through the reconnection region, then both the paths of the corresponding global manifolds and the lobe areas may change due to the reconnection. While this may alter the amount of flux in the turnstile lobes, the \emph{net} flux across the barrier cannot change if the associated fixed point field lines do not pass through the reconnection region.
\end{enumerate}

\section{Conclusion} \label{sec:conclusion}

We have proposed a method to define and measure reconnection in a 3D magnetic field stretching between two boundaries. The flux is partitioned using the global manifolds of hyperbolic fixed points of the field line mapping between the boundaries. Individual fluxes in the partition are defined as differences between the values of a generalized flux function ${\cal A}(x,y)$ at fixed points. This is a natural generalization of the flux function in a 2D magnetic field, and maintains the key advantage that the reconnection rate (with respect to this partition) is measured simply by the rate of change of ${\cal A}$ at fixed points. The associated partition reconnection rate is unique, and straightforward to compute: the main computational effort required is identifying the fixed points in the field line mapping at successive times.

\citet{petrisor2002} has previously recognised that reconnection may be related to values of an \emph{action} function (essentially ${\cal A}$) on hyperbolic orbits, though that analysis was limited to non-twist area-preserving maps. Our results are more general, and have been derived in a more physical way relating directly to the magnetic field itself. The interpretation of ${\cal A}$ as an action integral brings out a deep connection with the magnetic structure. \citet{cary1983} have shown that ${\cal A}(x,y)$ is the action in a variational formulation leading to the equations of the magnetic field lines. In other words, given ${\bf A}$, and assuming displacements $\delta{\bf x}$ with ${\bf A}\cdot\delta{\bf x}=0$ at the end-points, the Euler-Lagrange equations that extremise ${\cal A}$ are the equations of the field lines. 

Future work will consider in more detail how our partition reconnection rate compares with reconnection defined using the integrated parallel electric field, such as in numerical simulations (e.g. \citet{pontin2011}). There are certainly differences: we have seen in Section \ref{sec:example} how the visibility of a reconnection region (as defined with $E_{||}$) to our flux partition depends on its location. We interpret this as a difference in the \emph{topological effectiveness} of the reconnection, from the point of view of the flux partition. One way to refine the partition would be to integrate  ${\cal A}$ over more than one iteration of ${\bf F}_1$, measuring the values of ${\cal A}$ at the larger number of periodic points.

The example in Section \ref{sec:added} also demonstrated that the flux in each direction across a partial barrier may be much larger than the net flux, and furthermore may change under a non-ideal evolution even if the net flux remains constant. Our existing flux partition is insensitive to such changes. However, we note that it is possible to measure the flux of an individual lobe with the formula
\begin{equation}
\Phi(L) = \sum_{k=-\infty}^{\infty}\left[{\cal A}\big({\bf F}_1^k({\bf p}_2)\big) - {\cal A}\big({\bf F}_1^k({\bf p}_1)\big) \right],
\end{equation}
where ${\bf p}_2$ and ${\bf p}_1$ are the defining pips of the lobe. This may be proved using a similar geometrical argument to Theorem \ref{thm:netflux}, and is a straightforward generalization of a similar formula for lobe area in area-preserving maps \citep{mackay1984,mackay1986,easton1991}. Taking account of chaotic regions in this way is likely to be of particular importance in fields where heteroclinic tangles fill large areas of the plane, with multiple intersections. In this case, the structure of the field is dominated by chaotic regions. While the flux partition may still be defined in the same way using ${\cal A}$ values at the fixed points, the geometrical interpretation is less clear and requires further investigation.

Finally, a limitation of this work at present is our restriction to periodic fields where $B_z(x,y,1)=B_z(x,y,0)$. In future, we hope to relax this restriction thanks to recent theoretical progress in aperiodic dynamical systems \citep{malhotra1998}. The notion of hyperbolic trajectories, and their stable and unstable manifolds, may be extended to this more general setting.

\begin{acknowledgments}
This work was supported by the UK Science \& Technology Facilities Council, grant ST/G002436/1.
\end{acknowledgments}

% If in two-column mode, this environment will change to single-column format so that long equations can be displayed. 


%\begin{widetext}

%$$\mbox{put long equation here}$$

%\end{widetext}

% Figures should be put into the text as floats. 

% Use the graphics or graphicx packages (distributed with LaTeX2e).


%



% Put the label that you will use with \ref{} command in the braces of the \label{} command.

%

% \begin{figure}



% \end{figure}







%

% \begin{table}



% \end{tabular}

% \end{table}


%\begin{acknowledgments}

% Put your acknowledgments here.

%\end{acknowledgments}


\bibliography{fluxfunction}

\end{document}